\documentclass[dvips]{acta}
\usepackage{supertabular,lscape,epsfig}
\usepackage{amssymb}
\usepackage{amsmath}

\SetPages{179}{198}
 
\SetVol{61}{2011}

\begin{document}

\begin{Titlepage}
\Title{Oxygen issue in Core Collapse Supernovae.}
\Author{Abouazza Elmhamdi,}{Department of Physics and Astronomy, 
 King Saud University, PO Box 2455, Riyadh 11451, Saudi Arabia\\
 e-mail: elmhamdi@ksu.edu.sa}

\Received{ 2011}
\end{Titlepage}

\Abstract{
 We study the spectroscopic properties of a selected sample of 26 events 
 within Core Collapse Supernovae (CCSNe) family. Special attention is paid to 
 the nebular oxygen forbidden line [O I] 6300,6364\AA\ doublet. We analyze the 
 line flux ratio ``$F_{6300}/F_{6364}$'', and infer information
 about the optical depth evolution, densities, volume-filling factors
 in the oxygen emitting zones. The line luminosity is measured
 for the sample events and its evolution is discussed on the basis
 of the bolometric light curve properties in type II and in type Ib-c
 SNe. The luminosities are then translated 
 into oxygen abundances using two different methods. 
 The resulting oxygen amounts are combined with the recovered
  $^{56}$Ni masses and compared with theoretical models by means of
 the ``$[O/Fe]~.vs.~M_{ms}$'' diagram. Two distinguishable and continuous 
 populations, 
 corresponding to Ib-c and type II SNe, are found.  
 The higher mass nature of the ejecta 
 in type II objects is also imprinted on the 
 [Ca II] 7291,7324\AA\ over [O I] 6300,6364\AA\ luminosity ratios.  
 Our results may be used as
 input parameters for theoretical models studying the
  chemical enrichment of galaxies.}{Stars: Supernovae -- Core Collapse; Nebular spectra; Yields: Oxygen and Iron }

\section{Introduction}
 Recently, interest has
 increased the interpretation of Core Collapse 
 Supernovae (CCSNe) data, both spectra and photometry (Richardson et al. 2006; 
 Elmhamdi et al. 2006; Taubenberger et al. 2009; Maurer et al. 2010; 
 Elmhamdi et al. 2011). 
 In particular special
 attention has been devoted to the stripped-envelope events (i.e. type Ib-c
 hydrogen-deficient SNe). Studying enlarged samples of CCSNe 
 objects, having good quality observations, can be a potential tool 
 for assessing the similarities and the diversities within this SNe family, 
 relating these facts to the physics and possibly to the nature
 of their progenitors. 
 As an example, Elmhamdi et al. 2006 have presented an investigation
 of the spectroscopic properties of a selected optical photospheric
 spectra of CCSNe,
 discussing how hydrogen manifests its presence within this class. The authors
 argued for a low mass thin hydrogen layer with very high
 ejection velocities above the helium shell to be the most likely
 scenario for type Ib SNe.
 Although the primary goal of the the cited work was highlighting the hydrogen
 traces in CCSNe, an important by-product result concerns the behaviour of 
 oxygen, in particular the O I 7773 \AA\ line. Based on the 
 synthetic spectra fits, for this line it seems that at intermediate 
 photospheric phases, type Ib objects tend to have low 
 optical depths, while some type Ic SNe, e.g. SN Ic 1987M, are found 
 to display the deepest O I 7773 \AA\ profile. SNe of type IIb \& II, 
 at similar 
 phases, are found to be the objects with the lowest O I 7773 \AA\ optical 
 depth. 
 At somewhat later epochs, transient type Ib/c objects display deep 
 O I 7773 \AA\ troughs. Matheson et al. (2002) arrived at similar conclusions,
  indicating the O I 7773 \AA\ line to be stronger in SNe Ic than in SNe Ib. 

 Interestingly, the deeper and stronger
 permitted oxygen line O I 7773 \AA\ in photospheric spectra of SNe Ic and 
 Ib/c might imply that they are less diluted 
 by the presence of a helium envelope. We expect indeed the oxygen lines 
 to be more prominent for a ``naked'' C/O progenitor core. Two further
 observational aspects tend to reinforce this belief: first, the 
 forbidden lines, especially the [O I] 6300,6364 \AA\ doublet, seem to appear
 earlier following a SNe sequence ``Ic$-$Ib$-$IIb$-$II''. 
 The second indication comes from the fact that this nebular emission line 
 has a velocity width decreasing following the SNe sequence above. 
%%%%%%%%%%%%%%%%%%%%%%%%%%%%%%%%%%%%%%
\begin{figure}[htb]
\centerline{\psfig{file=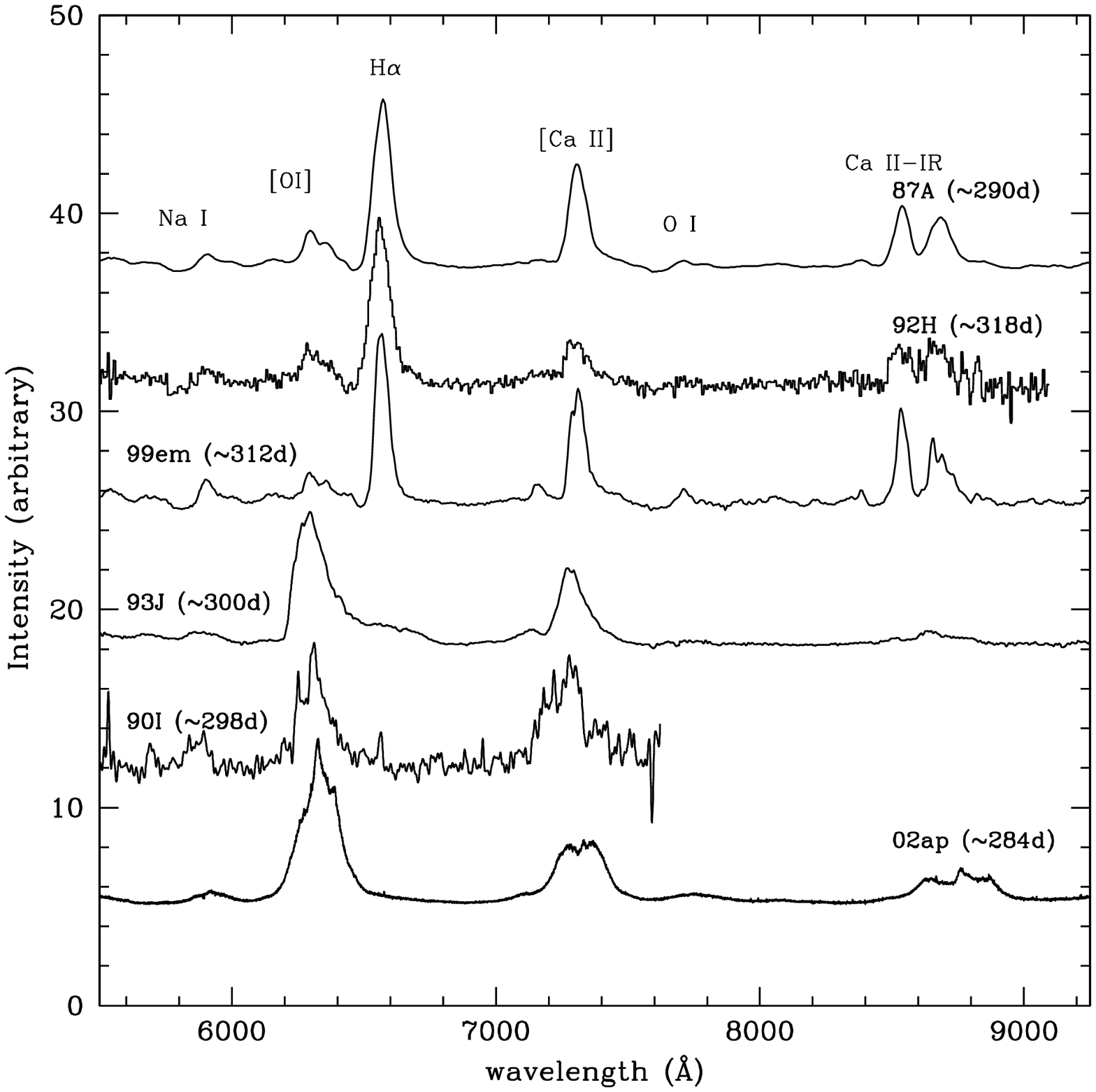,width=9.3cm,height=10cm}\psfig{file=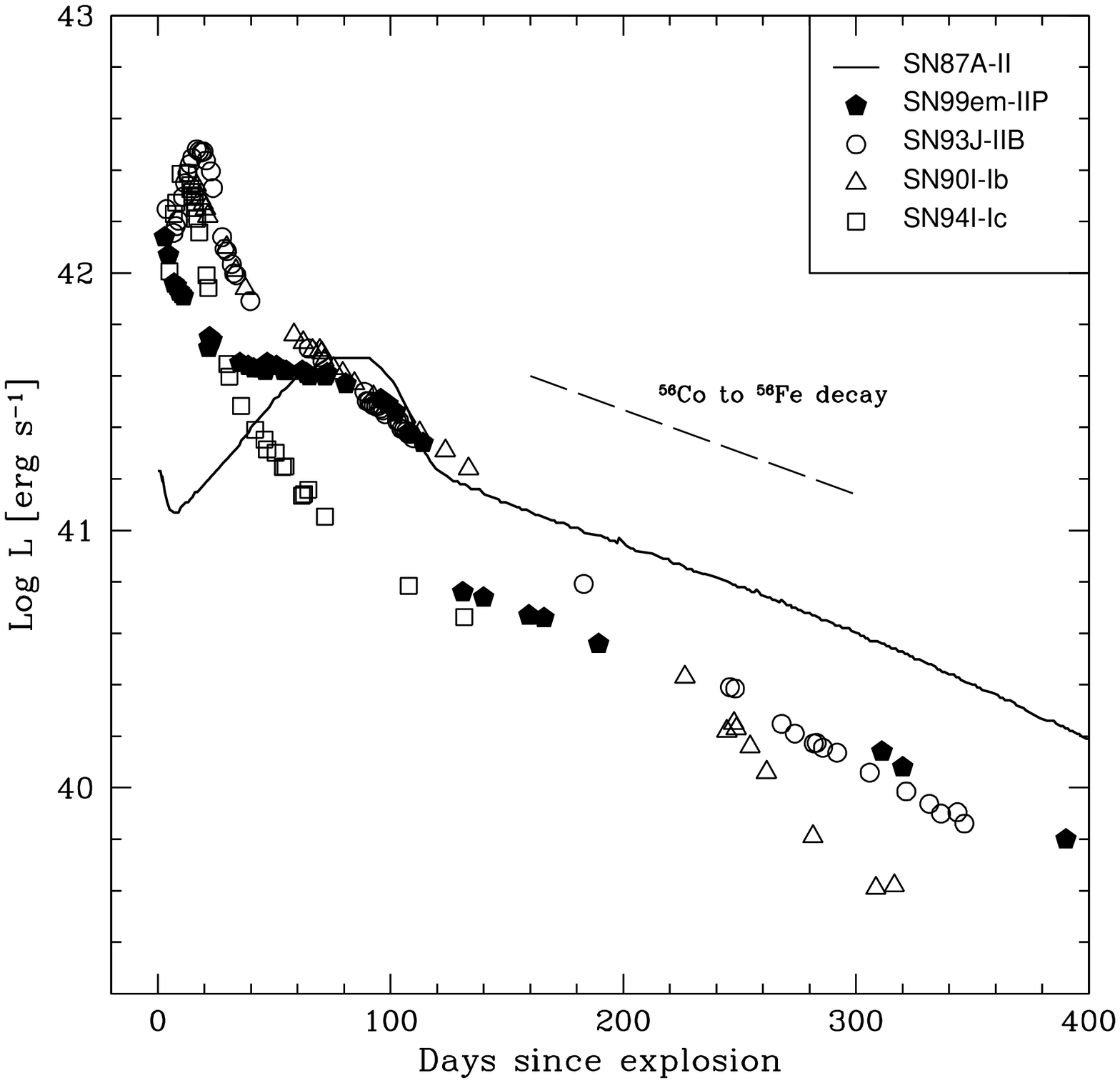,width=9.3cm,height=10cm}}
\FigCap{$Left~panel$: a sample of the late time CCSNe spectra. The most
 prominent lines are labeled. The corresponding observation date, since 
 explosion time, are also reported. $Right~panel$: the computed
 quasi-bolometric 
 light curve of SN II 1987A (UBVRI bands); SN IIP 1999em (UBVRI-bands);
 SN Ib 1990I (BVRI-bands); SN IIb 1993J (UBVRI bands) and SN Ic 
 1994I (BVRI bands). The $^{56}$Co to $^{56}$Fe decay slope is also shown
 for comparison.} 
\end{figure}
%%%%%%%%%%%%%%%%%%%%%%%%%%%%%%%%%%%%%%

 In the present work we explore these points and other
 issues related to oxygen lines within the CCSNe family,
  especially at late nebular phases. We analyze a sample of 26 CCSNe 
 events, and quantify the [O I] 6300,6364 \AA\ luminosities and 
  discuss their evolution in time. 
 The [Ca II] 7291,7324 \AA\ over [O I] 6300,6364 \AA\ flux ratio is also 
 shown and discussed 
 as a possible progenitor mass indicator (Fransson $\&$ Chevalier 1989).
 Using the the computed luminosities we discuss two methods to translate
 these measurements into masses. The estimated oxygen masses are combined
 then with the measured nickel masses, determined from
 light curves, and compared with yields from three known theoretical CCSNe
 models by means of the oxygen-to-iron mass ratio. This is potentially  
 important since our measurements can be directly correlated
 to the progenitor masses from theoretical models.    

 Worth noting here is the importance of oxygen and iron estimates, from 
 supernovae explosions, in the chemical enrichment and evolution
 of galaxies. In particular, the oxygen abundance is crucial in metal-poor 
 stars, and is a key issue in modeling the early phases of 
 the chemical galaxy evolution and as well in constraining the age of globular 
 clusters (VandenBerg et al. 2000; Melendez et al. 2001). 
 Oxygen is indeed considered as a major tracer of chemical 
 evolution, since it is one of primary elements ejected by CCSNe
 (i.e. resulting from massive stars). For example the oxygen-to-iron
 ratio, when [O/Fe]$>$0 such as in the halo of our Galaxy, is 
 an indicator of early chemical enrichment by massive stars.

 The paper is organized as follows.
 In Section 2, we describe the sample and we highlight the main
 spectroscopic characteristics and differences within the CCSNe family. Some 
 constructed quasi-bolometric light curves are also shown and discussed.
 The [O I] 6300,6364 \AA\ line luminosity and the 
 $F_{6300}/F_{6364}$ flux ratio are measured and presented in Section 3. 
 In Section 4, 
 we discuss two different methods for estimating oxygen mass in CCSNe.
 The methodology of $^{56}$Ni mass estimation is given. We also 
 evaluate the integrated flux ratio of the forbidden emission lines 
 [Ca II] 7291,7324 \AA\ and [O I] 6300,6364 \AA. 
  We conclude with a summary and discussion of our findings in Section 5.  
%%%%%%%%%%%%%%%%%%%%%%%%
\section{The sample}
%%%%%%%%%%%%%%%%%%%%%%%%%%
 The selected sample consists of 26 CCSNe objects$-$13 of them are type
 II, one of type IIb and 12 of type Ib-c. Data are gathered mainly
 from the literature (i.e. published available data).
 Use is made of the ``SUSPECT''\footnote{http://bruford.nhn.ou.edu/$\sim$suspect/
 index1.html} and of the 
``CfA''\footnote{http://cfa-www.harvard.edu/oir/Research/supernova
 /SNarchive.html} Supernova Spectra Archives. Some measurements
 are made on late spectra of SN 1996aq, taken from the 
 Padova-Asiago supernovae database.
 A summary of references and descriptive parameters of individual objects 
 is given in Table 1. 

 In what follows we adopt the standard reddening laws of 
 Cardelli et al. (1989).

 An example of the analyzed late spectra is shown in Fig.1(left panel), 
 together with identifications of the most prominent lines. We focus
 on the wavelength range 5500-9200 \AA.  
 At this epoch, when the events are in the radioactive tail phase, and except
 for the notable H$\alpha$ emission in type II SNe, the
 optical spectra are dominated by emission lines of 
 [O I] 6300,6364 \AA, [Ca II] 7291,7324 \AA\  and Ca II-IR triplet. The emission
 centered at $\lambda \sim$7800 \AA\ is usually attributed to O I 7774 \AA. 
  These aspects will be highlighted when evaluating the [O I]/[Ca II] line
 intensity ratio of the sample objects.
  
 Worth recalling here that the [O I] 6300,6364 \AA\ line
 is found to be absent or very weak 
 in SNe of type IIn\footnote{``n'' stands for 
 narrow. Type IIn are characterized by a narrow H$\alpha$ emission and
 high bolometric light curve with a relatively flat evolution.
 In the prototype SN IIn 1988Z indeed there was no sign of the nebular 
 forbidden lines [O I] 6300,6364 \AA\  and 
 [Ca II] 7291,7324 \AA. Similar behaviour is observed in SN IIn 1994aj 
 (Benetti et al. 1998). The well studied events SN IIn 1995N 
 (Fransson et al. 2002; Pastorello et al. 2005) and SN IIN 1995G 
 (Chugai $\&$ Danziger 2003; Pastorello et al. 2002) display the
  same characterizing peculiarity. 
 One possibility for this is that their progenitors are not massive,
  believed to explode in a very dense CS environment.}.
 
 In Fig.1 (right panel) the constructed ``quasi-bolometric'' light curves of 
 SNIIP 1999em, SNIIb 1993J, SNIb 1990I and SNIc 1994I are
 displayed. The peculiar SNII 1987A is also included for comparison.
 The available optical ``U,B,V,R,I'' broad-band photometric data have 
 been integrated to recover the pseudo bolometric light curves. 
 Literature references 
 for the photometry, together with the adopted 
 parameters are reported in Table 1. We do not include the IR-contribution
 since it is available only for SN 1987A, and hence the derived integrated
 bolometric light curves represent a lower limit to the ``real'' bolometric
 light curves. 
 The $^{56}$Co to $^{56}$Fe decay slope, $e-$folding time of 111.3 days, 
 corresponding to the full $\gamma -$ray trapping is also shown.

 The difference in the early CCSNe bolometric light curves, i.e. 
 the peak in type Ib-c
 SNe and the plateau in type II SNe, is mainly related to differences
  in presupernovae radii and structures. The plateau behaviour in type II
 is indicative of massive hydrogen envelopes, although its properties 
 (duration and luminosity) predict also a dependence on radius, energy 
 and the ejected amount of $^{56}$Ni (Popov 1993; Elmhamdi et al. 2003). 
 The lack of such significant hydrogen in the outer layers of type
 Ib-c inhibits the presence of a plateau behaviour.

 The peak characteristics, luminosity and width, are sensitive to the ejecta
 mass, released energy and the $^{56}$Ni mass (Arnett 1982).

 The clear faster decline at late phases for type Ib-c SNe is naturally 
 attributed to the significant $\gamma$-ray escape with decreased
 deposition as a result of low mass ejecta in this class of objects, while
 owing to the massive hydrogen mantle, type II SNe light curves indicate
 that radioactive decay of $^{56}$Co with the consequent trapping of
 $\gamma$-rays is the main source of energy powering the light curves at   
 late times. We note that in type II objects the V-light curve 
 follows the bolometric light curve 
 fairly well at late epochs, which simplifies for
 example the derivation of the synthesized $^{56}$Ni mass. It is
 sufficient then the use of the tail absolute V-light curve of SN 1987A
 , for which the ejected $^{56}$Ni mass is accurately known from 
 observations and detailed
 modeling, as a template for $^{56}$Ni mass derivations in other II events
 (Elmhamdi et al. 2003b; Hamuy 2003).
 The stripped-envelope SNe case is more difficult. The V-band light curve
 does not parallel the bolometric one. 
 It is hence necessary in type Ib-c a bolometric light 
 curve modeling rather a simple use of the absolute visual bands.     
%%%%%%%%%%%%%%%%%%%%%%%%%%%%%%%%%%%%%%%%%%
\begin{figure}[htb]
\psfig{file=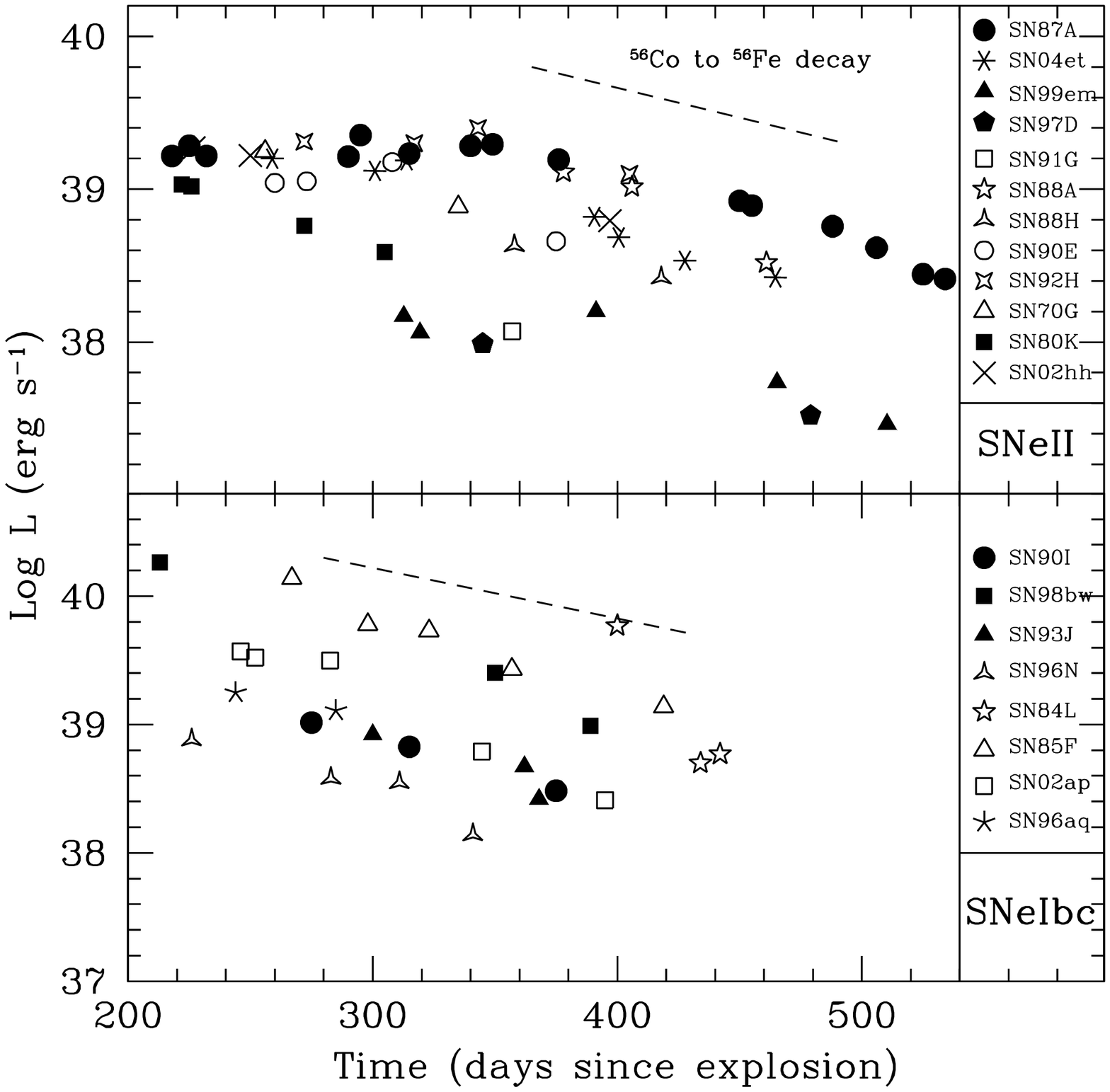,width=13cm,height=14cm}\FigCap{The temporal evolution of the [O I] 6300,6364 \AA ~luminosity for
 SNe of type II (upper panel) and for SNe Ib-c (lower panel). The $^{56}$Co
 to $^{56}$Fe decay slope is also reported.} 
\end{figure}
%%%%%%%%%%%%%%%%%%%%%%%%%%%%%%%%%%%%%%

\section{The analysis}
%%%%%%%%%%%%%%%%%%%%%%
 In this section we measure and study the oxygen [O I] 6300,6364 \AA\ line 
 luminosity for the CCSNe sample. For this purpose
 the available spectra were corrected for redshift and reddening effects 
 and calibrated with photometry when needed. 
 The recovered integrated 
 line fluxes, assuming a continuum level, are then converted to luminosities 
 using adopted distances.
%%%%%%%%%%%%%%%%%%%%%%%%%
%%%%%%%%%%%%%%%%%%%%%%%%%
\subsection{Oxygen Luminosity and Mass} 
%%%%%%%%%%%%%%%%%%%%%%%%%%%
\subsubsection{The luminosity}
%%%%%%%%%%%%%%%%%%%%%%%%%%% 
 Figure 2 highlights the [O I] 6300,6364 \AA\ line luminosity temporal 
 evolution, starting at 200 days after explosion, for type II SNe 
 (upper panel) and for type Ib-c SNe (lower panel). For comparison,
 the $^{56}$Co to $^{56}$Fe decay slope is also displayed (for an
 arbitrary $^{56}$Ni mass; dashed line).

 The emission [O I] 6300,6364 \AA\ light curves behave differently 
 within the CCSNe family. For type II events: the light curves have 
 a plateau-like maximum at day 200, changing slowly until day $\sim$340.
 At this epoch the light curves are already on the exponential decline 
 phase (Fig.1-right panel).
 The line-luminosity dropped then sharply, with a rate of decrease
 similar to that of the radioactive decay of $^{56}$Co 
 (i.e. $e-$folding time of 111.3 days). We note here
 that in the case of SN 1987A, it has been argued that dust condensation
 affects the line luminosity evolution starting
 at day $\sim$530, inducing a further increase in the rate of decrease
 (Danziger et al. 1991). At approximately the same time the [O I]
 line profile showed a marked shift of the peaks towards 
 blue wavelengths (Lucy et al. 1989). There are however two exceptions
 decaying earlier compared to the rest of the II SNe sample, namely
 SN 1970G and SN 1980K. Interestingly these two objects are
 respectively type IIP-L and IIL SNe. 
 For type Ib-c events: starting at the age of 200 days, the light curves
 are already on a steep decline, faster to that of the $^{56}$Co radioactive 
 decay. The ``Chi Square'' fit to SN Ib 1985F luminosity data for example
 indicates an 
 $e-$folding time of about 70 days.
 The time at which the deviation to the decline occurs varies
 among the CCSNe family, being earlier in Ib-c, followed by IIL and then later
 on SNe IIP. Specifically the line luminosity trend is 
 found to follow the bolometric light curves in the time
 range of interest, namely after 200d in type Ib-c and  after $\sim 340$d
 in type II. In SN 1987A for example, the [O I] luminosity
 relative to the bolometric one, i.e. $(L_{[O I]}/L_{bol}$), shows
 an almost flat-topped behaviour at the time interval $\sim 340-500$d. 
 Other effects enter at 
 later epochs, $>$ 500d, such as dust condensation and the 
 IR-catastrophe (Danziger et al. 1991; Fransson et al. 1996; Menzies 1991).
 On the one hand
 this is a direct evidence that the dominant source of ionization and 
 heating is $\gamma-$rays from the radioactive
 decay of $^{56}$Co in the CCSNe variety, with the $\gamma-$rays escaping with
 decreased deposition in type Ib-c events, owing to the low
 mass nature of their ejecta. 
 We note also that in both panels the line light curves span more than 
 one dex in luminosity, which may be related to the variation in the
 oxygen yields. Thus the importance of the oxygen mass estimates.
%%%%%%%%%%%%%%%%%%%%%%%%%%%%%%%%%%%%%%%
\begin{figure}[htb]
\centerline{\psfig{file=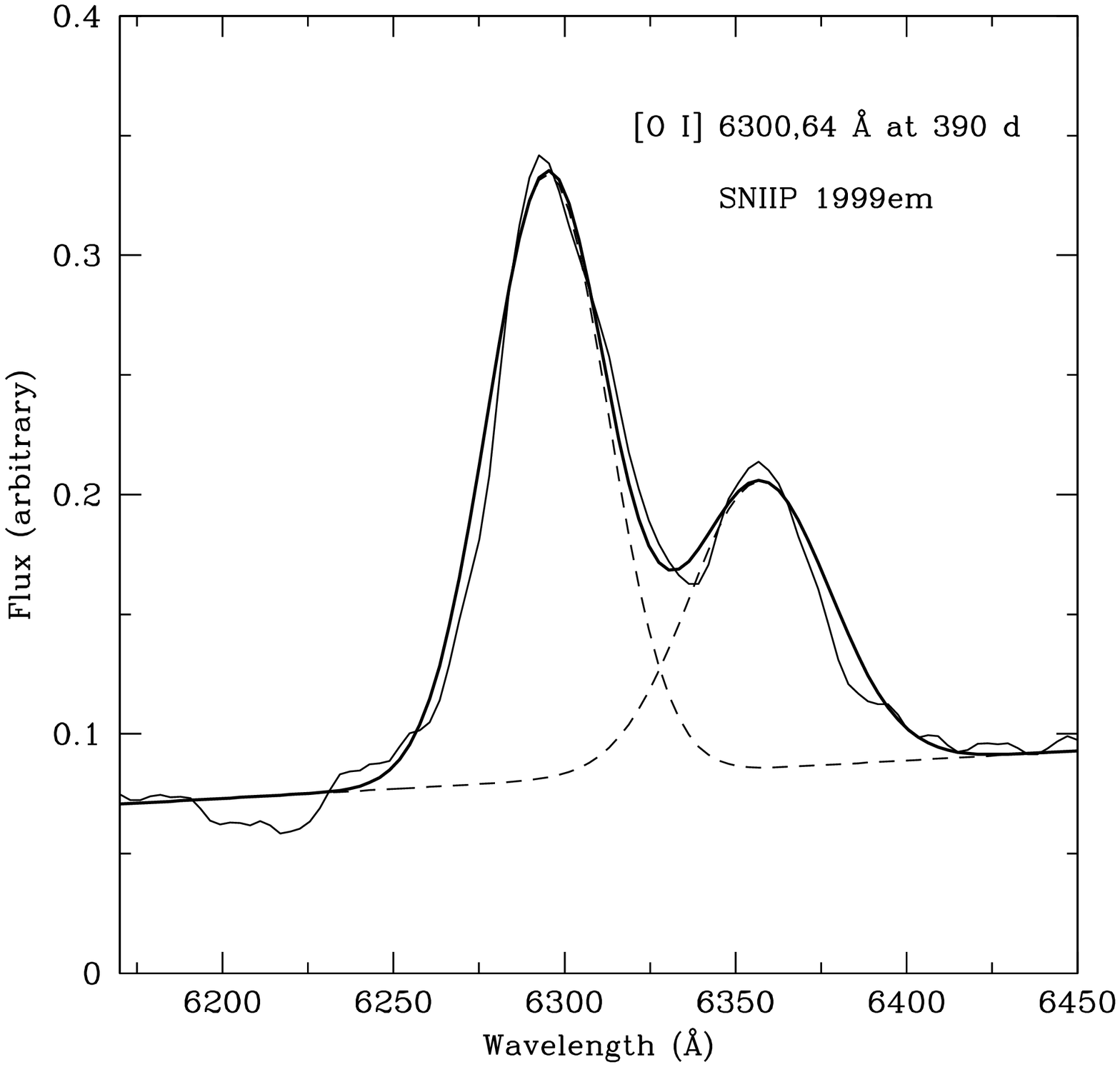,width=10cm,height=10cm}}
\FigCap{The fit example to the [O I] 6300,6364 \AA ~profile
 of SN IIP 1999em at day 390. Shown in thick line is the best fit.
 Dashed profiles refer to the decomposition of the best fit into
 two Gaussians for the [O I] 6300 \AA\ and [O I] 6364 \AA\ components.} 
\end{figure}
%%%%%%%%%%%%%%%%%%%%%%%%%%%%%%%%%%%%%%
%%%%%%%%%%%%%%%%%%%%%%%%%
\subsubsection{The flux ratio ``$F_{6300}/F_{6364}$'' } 
%%%%%%%%%%%%%%%%%%%%%%%%%
 Various investigators have dealt with the doublet flux ratio in 
 the [O I] 6300,6364 \AA\ line, especially for the well observed
  SN 1987A (Spyromilio $\&$ Pinto 1991; McCray 1996).
  It is well established that for a large and homologously expanding 
 atmospheres such as in supernovae 
 events, the Sobolev approximation holds and consequently simplifies 
 the radiation transfer and line formation treatments
  (Castor 1970; Jeffery $\&$ Branch 1990).   

 Within this context, a line intensity is given by: 
 $Flux \propto P_{esc}\times A_{ul}\times N_u $, where $N_u$ is the number 
 density
 of atoms in the upper state, $A_{ul}$ is the Einstein 
 coefficient, and $P_{esc}$ being the escape probability giving by
 $P_{esc}=[1-exp(-\tau)]/\tau$; $\tau$ is the Sobolev optical depth.
 For a freely expanding atmosphere the density
 $N_u$ goes as $t^{-3}$ and hence the Sobolev optical depth 
 should behave as  $\tau \propto N_u \times t=t^{-2}$.
 
 For the line of interest, i.e. the [O I] 6300,6364 \AA\ doublet, the
 two transitions have  $A_{6300}=3\times A_{6364}$,  and 
 $\tau_{6300}=3\times \tau_{6364}$; therefore the doublet 
 flux ratio reads:     

 \begin{equation}
  \frac{F_{6300}}{F_{6364}} = \frac{1-exp(-\tau_{6300})}{1-exp(-\tau_{6364})}
  = \frac{1-exp(-\tau_{6300})}{1-exp(-\tau_{6300}/3)}
 \end{equation} 

 A measure of the flux doublet ratio can be then translated into the
 optical depth in the 6300 \AA\ line in the SNe ejecta through solving
 Eq. 1. In addition we expect an asymptotic value for the ratio to be 1
 for the optically thick transitions and to be 3 for the optically
 thin case.

 Recovering the ratio of the flux doublet is not always an 
 easy task. Line blending can prevent an adequate intensity measurement.
 One possibility is a simple use
 of the peak intensities in the components of the doublet
 assuming it similar to the ratio of the flux of the doublet components 
  (Li $\&$ McCray 1992).
 Deblending yields to more accurate line ratio determination, especially 
 in cases with a sever blend. We selected a variety of  
 objects within our CCSNe sample that allow good estimates of the ratio.
 The studied events are: SNe IIP(1988A, 1999em, 2004et), SNe Ib(1985F, 1990I)
 and SNe Ic(1998bw, 2002ap). Data for SN 1987A are also included for
 comparison.  An example is illustrated in Fig.3 for the   
 observed spectrum of SN IIP 1999em at day 390. Our adopted 
 methodology consists in fitting the [O I] 6300,6364 \AA\ feature 
 with a single function formed by 2 Gaussians, 
  after estimating the continuum level and fixing the wavelength 
 separation of the doublet, i.e. 64\AA, and with 
 both Gaussians having the same velocity width. The fit in Fig.3 is 
 a good one. In some cases
 it was not possible to obtain a particularly good overall fit using only two
 Gaussians since one is left with either a residual in the blue wing of 
 the 6300 \AA\ component for narrower lines, or little evidence of the 
 the 6364 \AA\ component for broader lines. Elmhamdi et al. 2004 have used
 similar methodology and 
 obtained a satisfactory fit in the spectrum of SN Ib 1990I at 254d introducing
 a third Gaussian component assuming there is a contribution from 
 Fe II 6239 \AA\ emission, and allowing its velocity width to be a free 
 parameter (their Fig. 7).
 %Still to note here that a
 According to our sample analysis the two approaches, using the deblending 
 technique and a simple use of the peak fluxes, when possible, agree 
 within 20$\%$.   
 %%%%%%%%%%%%%%%%%%%%%%%%%
\begin{figure}[htb]
\centerline{\psfig{file=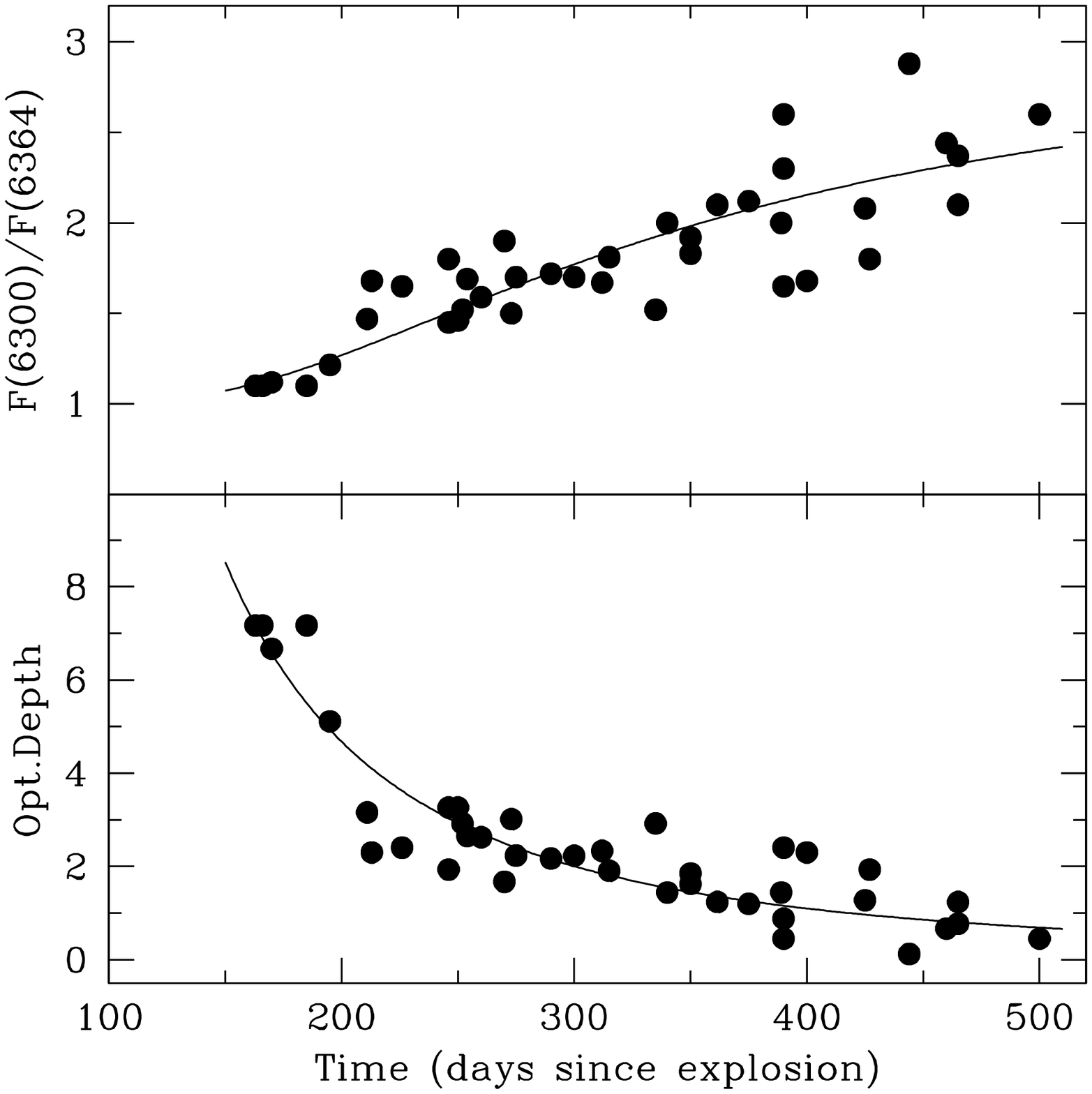,width=11cm,height=12cm}}
\FigCap{$Upper~ panel$: the temporal evolution of the flux ratio 
 $F_{6300}/F_{6364}$. The overploted curve corresponds to the 
 LSQ-fit by Eq. 2, indicating a value of $t_{trans} \simeq 426$d (see 
 text for details).
 $Lower~ panel$: the recovered optical depths in the [O I] 6300\AA\ line. 
 The power law LSQ-fit is also displayed ($\sim t^{-n}$ with $n= 2.089 
 \pm 0.153$; see text).} 
\end{figure}
%%%%%%%%%%%%%%%%%%%%%%%%%%%%%%%%%%%%%%
 Figure 4 reports the results of our investigation. The upper panel displays
 the temporal evolution of the ratio measurements, while in the lower panel
  we plot the corresponding optical depths in the primary component, 
 computed by solving Eq. 1.  
 The ratio values demonstrate a temporal trend
 from an optically thick phase to an optically thin one. The optical depth
 is found to decrease from values as high as $\sim 7$ around day 170
 falling to lower values at later phases ($< 0.5$ around 500d). The 
 continuous line
 is the resulted LSQ fit to the data, using a power-law function 
 (i.e. $\propto t^{-n}$). The best fit gives an index of 
 n= 2.089  ($\pm 0.153$). This is consistent with the index expected 
 from the simple expansion assumption discussed above 
 (i.e.  $\tau \propto N_u \times t=t^{-2}$).

 Furthermore, we may introduce a time dubbed $t_{trans}$, as the time at which
 the line makes the transition to the phase characterized by 
 $\tau_{6300} = 1$. In fact using $\tau \propto t^{-2}$
 and introducing $t_{trans}$, Eq. 1 can be written as:   
\begin{equation}
\frac{F_{6300}}{F_{6364}} = \frac{1-exp(-(t_{trans}/t)^2)}{1-exp(-(t_{trans}/\sqrt{3} t)^2)}
\end{equation}
Provided the ratio measurements, we fit the data with Eq.2. The LSQ fit is
 shown by the continuous line in the upper panel of Fig. 4. The best fit
 indicates a time of  $t_{trans} =426.06 \pm 13.24$ d. 
 The time at which $\tau_{6364} = 1$, is $245.96 \pm 7.64$ d.

 This above described analysis demonstrates the consistency within the CCSNe 
 family on how the components intensity ratio and the optical depths 
 develop in time.
%%%%%%%%%%%%%%%%%%%%%%%%%%%
\subsubsection{The mass} 
%%%%%%%%%%%%%%%%%%%%%%%%%%
In this section we adopt two different methods for estimating
the oxygen mass in SNe of type II and SNe of type Ib-c.\\ \\
{$\bullet$ \it Type II SNe:}
 
We use the recovered [O I] doublet luminosities to determine
 the abundance of the oxygen produced through the SN explosion.  
 
 At the epoch of about
 $1$ year, the luminosity of the [O I] doublet is powered by
 the $\gamma$-ray deposition and by
  ultraviolet emission arising from the deposition 
  of $\gamma$-rays in oxygen material. The [O I] doublet luminosity 
 is related directly to the oxygen mass (Elmhamdi et al. 2003a),
  and at a given time one may write: 
 \begin{equation}
L_t(\mbox{[O I]})=\eta \times  L_t(^{56}\mbox{Co}) \times \frac{M_{\rm O}}{M_{\rm ex}}
\end{equation}  
where $M_{\rm O}$ is the mass of oxygen,
$M_{\rm ex}$ is the $``$excited" mass in which the bulk of 
the radioactive energy is deposited, and $\eta$ is the efficiency 
of transformation of the energy deposited in oxygen into 
the [O I] doublet radiation. The L$_t(^{56}\mbox{Co})$ refers to 
 the radioactive decay energy input, given by 
 $L_t=L_{0}\times$($M_{\rm Ni}$/$M_{\odot}$)e$^{-t/\tau_{\rm Co}}$,  
 with $L_{0}\simeq 1.32 \times$ 10$^{43}$ ergs s$^{-1}$, 
 the initial luminosity corresponding to 1 $M_{\odot}$  
 and $\tau_{\rm Co}=111.3$ d.

  Assuming then that all type II events have similar $\eta$ and 
 $M_{\rm ex}$ at similar phases, we derive rough estimates
 of the oxygen mass for the events of the sample adopting [O I] 6300,6364\AA
 ~light curve of SN 1987A as template. As discussed previously, 
 the [O I] luminosity is found to have similar decay
 rates in SNe II at late phases, following a behaviour similar to 
 the bolometric light curves. This fact gives some confidence for the
 use of SN 1987A [O I] luminosity as a template for the recovery of the
 amounts of ejected oxygen. The oxygen mass in SN~1987A is 
 estimated to be in the range $1.5-2$ $M_{\odot}$ 
 (Fransson et al. 1996; Chugai 1994).

 In Table 1, column 7,  we report the amounts of the ejected oxygen mass 
 derived in this manner. The variation range, for each event,
 reflects the combination of the variation in the oxygen mass for 
 SN 1987A (i.e. $1.5-2$ $M_{\odot}$) together with the uncertainties
 from the fit procedure. 
\\ \\
{$\bullet$ \it Type Ib-c SNe:}

 The emission lines which are formed at densities above
 their critical ones, given the optical depths are not large, have 
 the luminosity directly proportional to the mass of the emitting ion.
 Such conditions hold for the [O I] 6300,6364 \AA\ doublet line emission
 in type Ib-c SNe at nebular phases. 
 On the one hand, we found moderate optical depths at late phases 
 (Fig 4, lower panel). On the other hand, the condition of   
  the high density limit above the critical density for the [O I] line 
 ($\sim7~\times 10^5 ~$cm$^{-3}$) is found to be clearly fulfilled in 
 the ejecta of type Ib-c SNe (Leibundgut et al. 1991; Elmhamdi et al. 2004). 
 A possible direct method can be used to check the high density limit  
 characteristic. Indeed, the density is directly related 
 to the relative 
 strengths of the [O I] doublet components (Leibundgut et al. 1991; 
 Spyromilio $\&$ Pinto. 1991). In both cited works, the variation of the line 
 doublet ratio as function of the density at a given time is computed, and is
 found to be insensitive to the adopted temperature, especially for 
 the late epochs.
 From our computations we take a ratio value of 1.58 at day 250 or 1.8 at day
 300 as representatives (i.e. from the best fit in Fig.4-upper panel). 
 According to Fig. 6 of Leibundgut et al. (1991), the 
 uncertainty in the temperature leads to the following density 
 range
 $2\times 10^{9}~$cm$^{-3}\leq N_{e} \leq 4\times 10^{9}~$cm$^{-3}$.

 In the high density limit, i.e. above the critical density, the mass 
 of ejected oxygen can be recovered using the 
 [O I] 6300,6364 \AA\ flux. Uomoto (1986) has shown that the oxygen mass,
  in M$_\odot$, is given by:

\begin{equation}
M_{Ox} = 10^{8} \times D^2 \times F(\rm{[O ~I]}) \times \it exp{(2.28/\it T_4)}
\end{equation} 
where $D$ is the distance to the supernova (in Mpc), $F$ is the reddening-free
 [O I] integrated flux (in ergs s$^{-1}$ cm$^{-2}$) and $T_4$ is the 
 temperature of the oxygen-emitting gas (in 10$^4$ K).

 Worth noting that because of the variation of $F$([O I]) and $T_4$, Eq. 4 
 implies time-dependence. 
 Schlegel $\&$ Kirshner (1989), when estimating the ejected
 oxygen amounts, have adopted a 
 constant temperature $T_4$=0.4 at the nebular phase of SNe Ib 1984L and
 1985F. The assumption of a 
 constant temperature at different late phases provides different oxygen 
 masses as one may expect, since earlier nebular epochs are hotter compared to 
 latter ones. 

 Alternatively, the temperature at a given time can be constrained
  based on the 
 [O I] 5577 \AA\  to [O I] 6300,6364 \AA\ flux ratio. Assuming 
 that the O I lines are formed mainly by collisional excitation, the ratio 
 is given by the following expression (Houck \& Fransson 1996):   

\begin{equation}
\frac{F_{6300}}{F_{5577}}=0.03 \beta_{6300} \times [1+1.44 \it T_3^{-0.034}(\frac{10^8}{N_e})]\times \it exp{(25.83/ \it T_3)}
\end{equation}
%\vspace{-0.3truecm}\hspace{2truecm}$\times \it exp{(25.83/ \it T_3)} $

\vspace{0.3truecm}\hspace{-0.4truecm}where $T_3$ is the temperature of 
 the oxygen-emitting gas (in 10$^3$ K), $\beta_{6300}$ is the 
 [O I] 6300 \AA\ Sobolev escape probability ($\simeq$ 1) and $N_e$ is the 
 electron density. 
%%%%%%%%%%%%%%%%%%%%%%%%%%%%%%%%%%%%%%
\begin{figure}[htb]
\centerline{\psfig{file=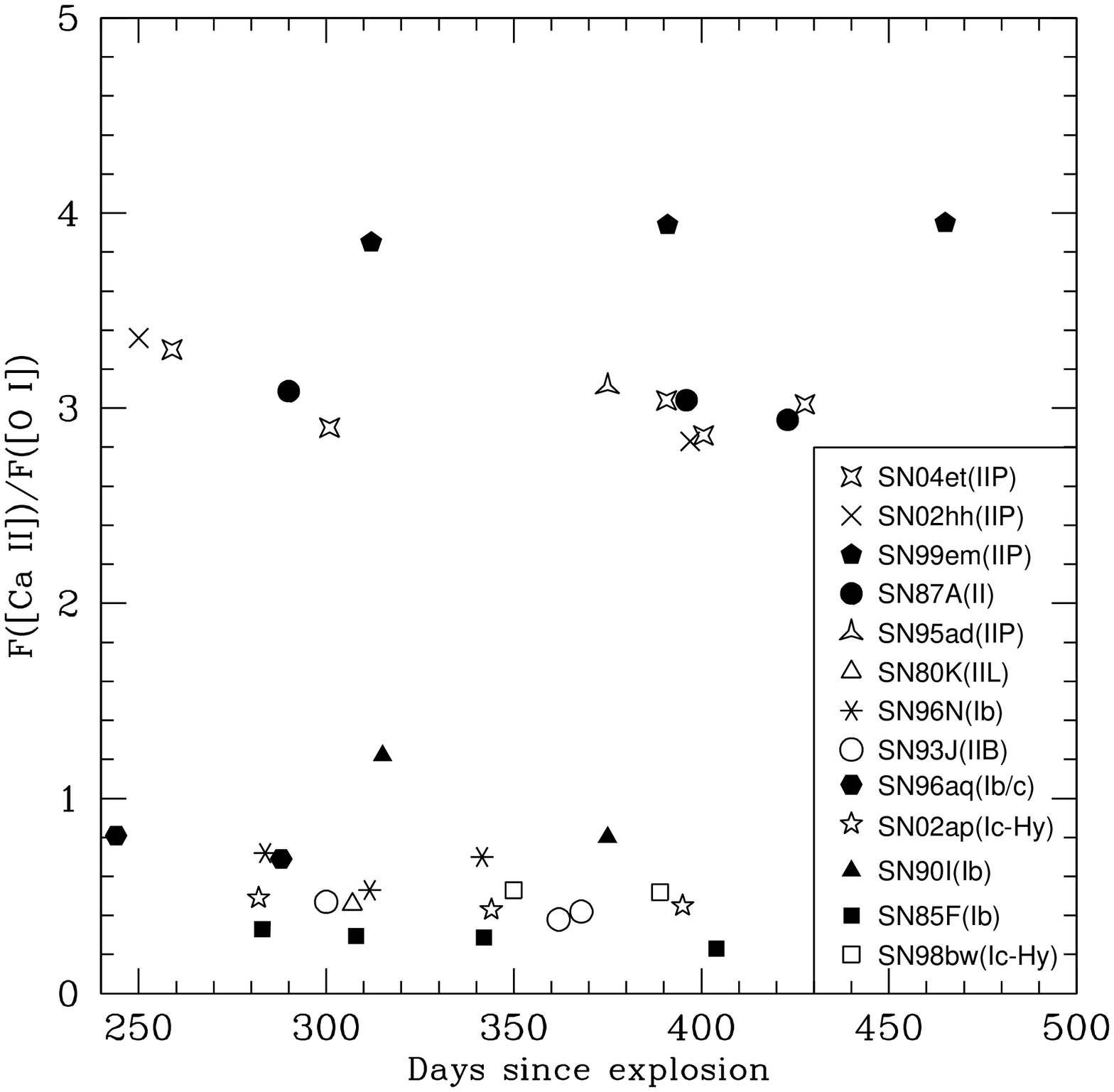,width=11cm,height=11cm}}
\FigCap{The temporal evolution of the [Ca II] 7291,7324 \AA\ over 
 [O I] 6300,6364 \AA\ integrated flux ratio. 
 Note the complete 
 separation of type Ib-c from type II SNe.}
\end{figure}
%%%%%%%%%%%%%%%%%%%%%%%%%

 A problem with this method is the very weak observed [O I] 5577 \AA\  feature 
 in the late time spectra analyzed here. It is indeed an indication of low 
 temperatures in the oxygen-emitting zone. However one may estimate an upper 
 limit on the [O I] 5577 \AA\ flux by integrating over the same velocity 
 interval of the [O I] 6300,6364 \AA\ profile. It is found that the 
 temperatures at late phases of interest, $\ge$250d, tend to have values
 in the range $3400 - 4200$K.

 Results from this described method, through equation 4, are listed in
 column 7 in Table 1. For SNe 1994I, 1993J, 1998bw and 2002ap however,
 the amounts reported in the table come from the most recent 
 spectroscopic and photometric modeling (see column 8 for references).

 \subsubsection{The [Ca II]/[O I] intensity ratio}  

  After considering the reddening correction, we have evaluated the 
  integrated flux ratio of the forbidden emission lines 
 [Ca II] 7291,7324 \AA\ and [O I] 6300,6364 \AA, for the CCSNe sample objects 
  having wavelength coverage in these two nebular lines. 
 Results are displayed in Fig. 5. Two separate 
 populations are clearly distinguishable. The ratio is found to remain
 below unity in type Ib-c, with the exception of SN 1990Ib around day 315
 ($\sim$1.22).
 SNe IIb 1993J, Ic-hypernova 1998bw and IIL 1980K belong to this category 
 as well. A mean value is measured to be $\sim$ 0.51.
 SNe of type II instead concentrate on the top of the figure, with a mean 
 value of $\sim 3.17$. Additionally, the two classes appears to have
 a flat evolution behaviour of the [Ca II]/[O I] ratio.
  
 In a detailed analysis, Fransson $\&$ Chevalier (1987, 1989) have modeled
 late emission spectral lines for different supernova progenitors. It is
 found that because of the composition and density structures one can
 use the relative emission line strengths as a progenitor indicator tool.
 In particular the forbidden emission line ratio [Ca II]/[O I] presents 
  weak dependence on density and temperature of the emitting zones, and
 is expected to display an almost constant evolution at late epochs. The ratio
 is found to be very sensitive to the core structure and mass. Furthermore
 it seems that the ratio tends to increase with decreasing progenitor
 mass. The distribution of our measurements in two groups tend to indicate
 different progenitor properties, with lower progenitors for type Ib-c, IIb
 and IIL classes compared to normal type IIP events. It is worth to note here
 however that in type Ib-c SNe there is no hydrogen rich Ca II emitting 
 zone as is the case for type II objects (de Kool et al. 1998).
  
 \subsubsection{The $^{56}$Ni mass} 

 As discussed in Sect. 2, the exponential behaviour of the late V-band absolute
 light curves of type II SNe is found to be in accordance
 with the radioactive decay $^{56}$Co$-$$^{56}$Fe ($M_V \propto 
 exp(-t/111.3~ $d$)$), which provides a potential
 method to recover the amount of the ejected $^{56}$Ni with the use of the  
  $M_V$ photometry of SN 1987A as a template between $120-400$ d 
 (Elmhamdi et al. 2004; Hamuy 2003). An amount of $0.075~M_{\odot}$ is
 adopted for the $^{56}$Ni mass of SN 1987A (Catchpole et al. 1988, 
 Suntzeff \& Bouchet 1991). Results from the described methodology are 
 reported in Tab. 1 (column 6). 
 There is a significant scatter of the 
 ejected $^{56}$Ni masses, with an average value of $\approx 0.053~M_{\odot}$.

 For type Ib-c events, the broad band light curves do not trace necessary
 the bolometrics. The amounts summarized in Tab. 1(column 6) are indeed results
 from bolometric light curves and/or spectra modeling (see column 8 for
 the corresponding references). An average value of 
 $\approx 0.18~M_{\odot}$ is computed.   
 
\section{Discussion and conclusions}

 We have selected a sample of 26 events within the CCSNe family.
 Our main goal was to investigate how the oxygen manifests its presence 
 at late phases for each SN class, especially the emission 
 [O I] 6300,6364 \AA\ doublet.
 
 Based on investigating early spectra of CCSNe, the 
 permitted oxygen line O I 7773 \AA\ seems to get weaker following the
 SNe sequence ``Ic$-$Ib$-$IIb$-$II''
 (Elmhamdi et al. 2006; Matheson et al. 2001).
 For stripped envelope objects, being less diluted by the presence
 of an helium envelope, one may expect the oxygen lines
 to be more prominent. An observational fact that clearly support
 this belief is the earlier appearance of the [O I] 6300,6364 \AA\ 
 following the above order, i.e. ``Ic$-$Ib$-$IIb$-$II''.          
 The following examples highlight this fact. The 
 oxygen line emerges at an age of 1-2 months in
 type Ic SN 1987M (Filippenko 1997). SN Ic 1994I displayed 
 evidence for the line at an age
 of 50 days, although some hints may even be seen in the $\sim$36 days 
 spectrum (Clocchiatti et al. 1996b). SN Ic 1997B shows clear evidence
 for [O I] 6300,6364 \AA, [Ca II] 7291,7324 \AA\ and also O I 7773 \AA\ 
 on the 2 months spectrum (Gomez $\&$ Lopez 2002).
 In SN 1998bw, Ic-hypernova event, the nebular features were already
 recognizable in the 43 days spectrum (Patat et al. 2001). While
 in SN Ib 1990I it was hinted at the 70 days spectrum (Elmhamdi et al. 2004). 
 In other type Ib SNe it appears
 earlier than in SN 1990I. In SN IIb 1993J, a transition object, the line 
 was visible in the 62 days
 spectrum (Barbon et al. 1995). SN 1996cb, another well 
 observed IIb event, showed
 evidence of the [O I] 6300,6364 \AA\ line around day 80 (Qiu et al. 1999).
 In SNe II, however, the line appears later: around day 150 in SN 1987A 
 (Catchpole et al. 1988) and after day 138 in SN 1992H 
 (Clocchiatti et al. 1996a). In SN II 1999em it is suggested at a somewhat
 earlier phase compared to SNe 1987A and 1992H, namely at day 114. 

 Furthermore, the width velocities in the [O I] 6300,6364 \AA\  and 
 in the [Ca II] 7291,7324 \AA\  nebular lines are mainly found to be
 higher in SNe Ic than in SNe Ib (Schlegel $\&$ Kirshner 1989; 
 Matheson et al. 2001). 
 In the cited papers, the FWHM values were evaluated fitting a
 single profile, namely Gaussian, to the total observed profiles.
 We mention here that given the line blending in these two lines (doublets), 
 results should be interpreted with some caution. One should indeed resolve
 the lines and compare width velocities of single components rather
 than using a single Gaussian fit to the whole feature. It is not
 simple however in some cases to get a satisfactory fit, especially
 when the blend is sever. According to our analysis, we faced more
 difficulties in type Ib and Ic than in type II. This fact is indeed
 in itself indicative of higher velocities in stripped envelope objects, while
 in SNe II, owing to lower expansion velocities, the 
 observed [O I] 6300,6364 \AA\ profile
 is less complex and the two components are clearly visible.
 This can be easily understood if considering the formation of the 
 line doublets 
 in cases where the element expansion velocity does not exceed the
 the two components separation velocity, i.e. 3000 km s$^{-1}$.
 We have checked the width velocities by means of the FWHM of 
 the single [O I] component at 6300 \AA\ of some of our sample events.
 At a phase of $\sim$300d, SNe II 1987A and 1999em 
 are found to have respectively FWHM $([O~I]~6300~\AA)\sim 2750$ 
 and $2400$ km s$^{-1}$, while for Ib SNe a typical range variation of 
 $\sim 4000-5000$ km s$^{-1}$ is deduced (e.g. SNe Ib 1985F, 1996N). 
 SN IIb 1993J, at similar age, has a velocity of $\sim 4600$ km s$^{-1}$,
 with a complex [O I] 6300,6364 \AA\ profile due to the presence of 
 H$\alpha$ at the red wing of the line (Patat et al. 1995).
 Higher velocities, $\geq 6000$ km s$^{-1}$, are computed from Ic spectra 
 (e.g. SNe 1987M, 1994I and 1998bw).   
 The [O I] 6300,6364 \AA\ nebular line at late epochs is representative
 of the expansion velocity. For a given explosion energy, a 
 greater ejecta mass allows for lower velocities. The found lower
 velocities in type II SNe can be attributed then to a large
 mass with respect to type Ib-c objects. 

 We have investigated in details the properties of the 
 [O I] 6300,6364 \AA\ doublet profile. The measured line luminosities
 are found to trace well the bolometric
 light curves, behaving differently in type Ib-c than in type II events. 
 In SNe II, the line light curves show a plateau-like evolution 
 until $\sim340$ days, following later-on by an exponential decline similar to 
 the radioactive decay of $^{56}$Co ($e-$folding$\simeq$111.3 days). The
 linear type II SNe display an early decline in their [O I] light curves.
 type Ib-c light curves display already a decline by an age of 200 days, 
 steeper than the $^{56}$Co radioactive decay rate.
 The $\gamma-$rays from the radioactive decay of $^{56}$Co are hence
 the dominant source of ionization and heating  in CCSNe variety.
 Furthermore, owing to the low mass nature of their ejecta, type Ib-c SNe 
 are characterized by a decreasing deposition of the $\gamma-$rays escape. 
 Similar conclusions about the progenitor diversity are argued from
 the analysis of the integrated flux ratio of the forbidden emission 
 lines [Ca II] 7291,7324 \AA\ and [O I] 6300,6364 \AA. The ratio is
 found to concentrate in two distinguishable locations, namely around 
 a value of 3 for type II SNe (mean $\sim3.17$), and below unity 
 in Ib-c objects (mean $\sim0.51$). This forbidden lines ratio is potentially
 sensitive to the progenitor mass star (Fransson $\&$ Chevalier 1987; 1989), 
 suggesting higher masses in SNe of type II.
 
 The way CCSNe family members transit from optically thick to optically 
 thin phases is imprinted onto the profile of the [O I] 6300,6364 \AA\ doublet.
 We emphasize this point making use of the components flux 
 ratio $F_{6300}/F_{6364}$,
 after deconvolving the two components of the observed line profiles. 
 Based on a simple description of the physics behind the line formation, 
 our results indicate a consistency within the CCSNe 
 on how the components intensity ratio and consequently 
 the optical depths develop in time. Additionally, the 
 ratio $F_{6300}/F_{6364}$ can be used to estimate the average density 
 of the oxygen-emitting zone (Leibundgut et al. 1991; 
 Spyromilio $\&$ Pinto 1991).  
 Adopting representative values of the ratio of 
 1.58 at day 250 or 1.8 at day
 300, inferred from our best fit in Fig.4-upper panel, 
 and according to Fig. 6 of Leibundgut et al. (1991), the 
 uncertainty in the temperature leads to the following density 
 range $2\times 10^{9}~$cm$^{-3}\leq N_{e} \leq 4\times 10^{9}~$cm$^{-3}$.
 Worth noticing here that based on our approach one may get further 
  informations about the physical conditions in the oxygen emitting region.
 In fact, for an uniform density distribution, one may use the 
 above derived density range together with the element volume, based
 on the [O I] 6300 \AA\ FWHM velocity, to get a rough estimate of the
 ejected oxygen mass. This approach is found to provide too large
 amounts compared to results reported in Tab. 1 (column 7).
 For SN IIP 1999em for example, at 300 days, using the previous derived 
 velocity of 2400 km s$^{-1}$, we find an ejected oxygen 
 mass as high as $\sim$40 M$_\odot$, indicating a volume-filling factor
 as low as 10$^{-2}$. Similar filling factor is found for SN II 1987A and 
 SN Ib 1990I. These results suggest
 that the oxygen material in CCSNe generally fills its volume in a clumpy
 way rather than homogeneously.   
   
 We describe and adopt two methods for the oxygen mass estimate in CCSNe.
 We provide also estimates of the ejected $^{56}$Ni masses. 
 It should however be noted that various parameters affect the 
 recovered values, especially as long as the oxygen mass is concerned, 
 contributing hence to their uncertainties. Indeed, the luminosity and 
 temperature measurements are sensitive to the nature and quality of 
 the spectra (i.e. 
 severe line blending; multi-component profiles; presence of [O I] 5577 \AA).
 According to our test effects, we estimate the uncertainty of the 
 derived oxygen-mass amounts reported in Tab. 1 to vary within $15-30 \%$.   
 Furthermore, for cases with clear evidence of O I 7774 \AA\ line, the presence
 of ionized oxygen is argued (Mazzali et al. 2010), requiring higher mass of 
 oxygen than the one needed to produce [O I] 6300,6364 \AA\ alone. 
 Generally, Nickel masses are found to suffer less uncertainties 
 compared to oxygen.
 The errors in type II SNe, being more homogeneous than SNe Ib-c, do not 
 exceed $20 \%$ (mainly due to distance and reddening estimates). In SNe Ib-c,
 we estimate a variation of $10-30 \%$.''

 These estimates
 , i.e. oxygen and iron masses, are of importance as they can be
 indicative of the core mass and related to the progenitor star nature. 
 For the purpose of a better investigation, we compute the 
 [O/Fe]\footnote{$[A/B]=log_{10}(A/B)_{\star}-log_{10}(A/B)_{\odot}$} 
 yields ratio for each event. A solar value of 
 $~log_{10}(O/Fe)_{\odot} = 0.82$ 
 dex (Anders $\&$ Grevesse 1989) is adopted. The recovered 
 amounts are reported in Fig. 6 (horizontal lines) as function of the initial 
 mass according to the most reliable core collapse SN models in literature
 , namely Woosley $\&$ Weaver 1995 
 (short dot-dashed line), Thielemann, Nomoto $\&$ Hashimoto 1996 
 (long dashed line) and Nomoto et al. 1997 (long dot-dashed line). 
 A typical value for type Ia SNe is 
 as well reported (Nomoto et al. 1984), which shows the nature of type Ia 
 SNe being iron producers. 
 
 Fig. 6 defines two possible concentration regions,
 of type II objects from one side and type Ib-c from the other side, and 
 suggestive of a continuum in the $[O/Fe]$ values. 
 Indeed, an important issue that can be read out immediately from Fig. 6 is 
 that the reported results for type Ib-c are found to be located at the 
 bottom of type II SNe. 
 These zones may provide constraints on the progenitor masses 
 in CCSNe family. 
 Within the core collapse models of type Ib-c SNe, two scenarios are 
 argued: $first$ a single high mass star ($M_{ms} \geq 35~$ M$_\odot$) 
 exploding as Wolf-Rayet star after an episode of strong stellar wind, and 
 $second$ a less massive star ($M_{ms} \sim 13-18~$ M$_\odot$) in a binary 
 system. Although the progenitor nature of type Ib-c SNe is an open and debated
 issue, the position of type Ib-c events in the ``$[O/Fe]~.vs.~M_{ms}$'' 
 plot, according to Fig. 6, might be taken as an observational 
 support of the intermediate massive stars in binary systems, stripped of
 their envelope through binary transfer, as the favoured 
 progenitors in this class of objects.

 Although the uncertainties in the stellar evolution models of massive
 stars and in the determination of the oxygen and iron
 yields, Fig. 6 provides a methodology to elucidate 
 the progenitor nature of core collapse SNe and interesting comparisons
 can be drawn . However, extended samples and 
 more reliable determinations, especially for oxygen abundances (late spectra 
 modeling for example), are clearly 
 needed to populate the ``$[O/Fe]~.vs.~M_{ms}$'' diagram and hence have a
  deeper view on the oxygen to iron ratio and how it changes as function of 
 the type of SN. 
 The present quantitative analysis can also provide insights and input
 data for the evolution of galaxies and the chemical enrichment
 models. 

 The main goal of the paper was to highlight some late-phases spectra
 related properties of CCSNe, providing possible physical 
 interpretations. As a first step we concentrate our study on 26 selected 
 events.
 For future investigation we aim to enlarge the studied sample including
 more recent and up-dated observed CCSNe objects, especially those with 
 spectra of improved time coverage and spectral resolution 
 (mostly last decade observations; Elmhamdi et al. In preparation).   

\Acknow{We thank the referee for the very helpful and constructive suggestions.
 We thank I. John Danziger for his comments on the original
 manuscript and for the stimulating discussions.
We are grateful for the use of the ``SUSPECT'' SNe Archive-Oklahoma 
University and also of the ``CfA'' Supernova Archive.
We also thank D. K. Sahu for providing published data of SN IIP 2004et,
R. Foley for the discussion about SN Ic 2002ap data and M. Pozzo
for discussing SN IIP 2002hh. This project was supported by King Saud 
University , Deanship of Scientific Research, College of Science Research 
Center.}  
%\end{acknowledgements}

%%%%%%%%%%%%%%%%%%%%%%%%%

\newpage
%%%%%%%%%%%%%%%%%%%%%%%%%%%%%%%%%%%%%%
\begin{figure}[htb]
\centerline{\psfig{file=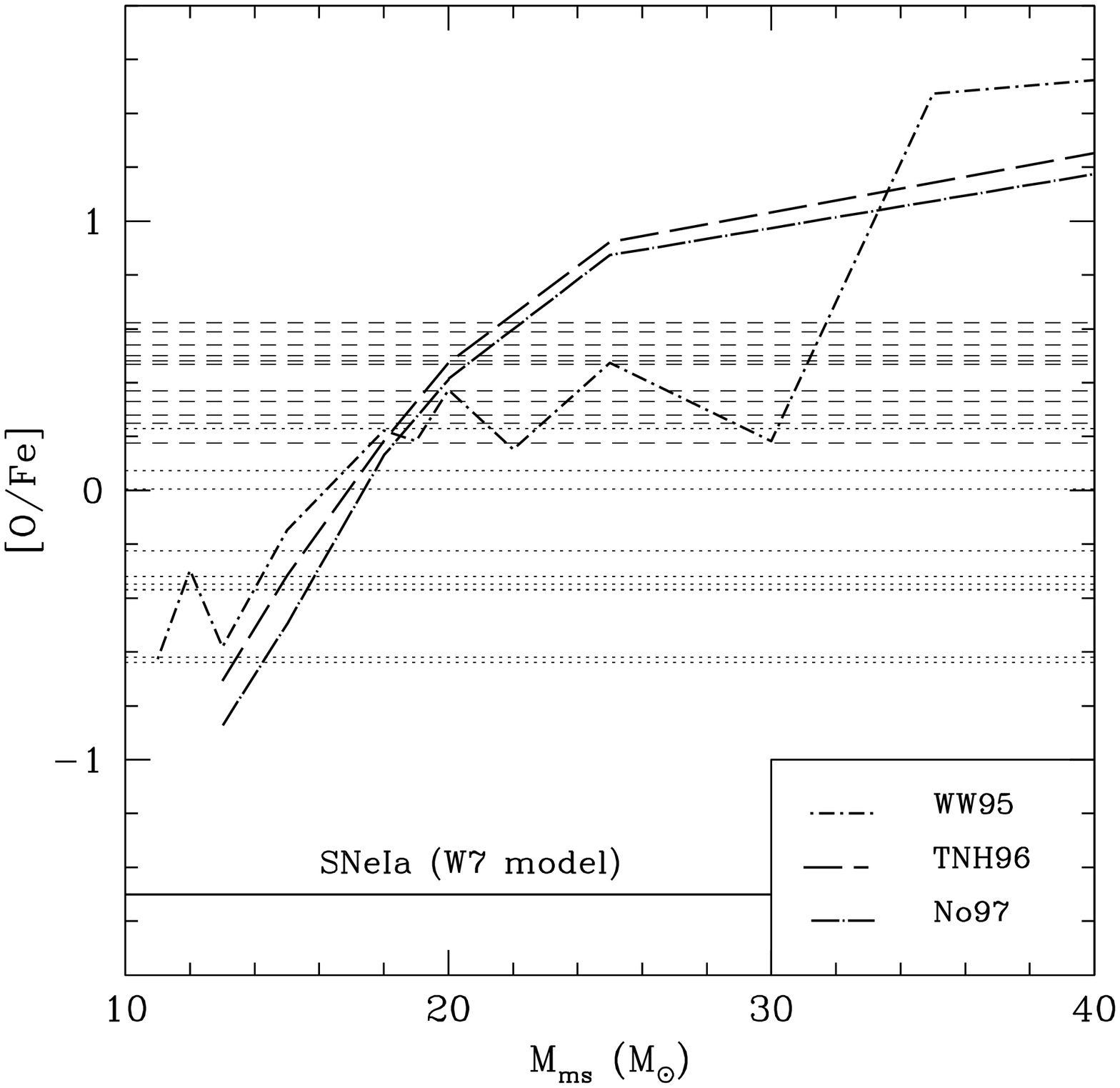,width=17cm,height=17cm}}
\FigCap{The [O/Fe] yields ratio as a function of the initial mass according to 
 the  theoretical models of Woosley $\&$ Weaver 1995 (short dot-dashed line),
  Thielemann, Nomoto $\&$ Hashimoto 1996 (long dashed line) and 
 Nomoto et al. 1997 (long dot-dashed line). For each event 
 the corresponding estimated [O/Fe] value is plotted horizontally (SNe II with
 horizontal dashed lines, while horizontal dotted lines correspond to 
 SNe Ib-c). 
 Typical value for 
 thermonuclear SNe is also reported for comparison (horizontal continuous 
 line, Nomoto et al. 1984). }
\end{figure}
%%%%%%%%%%%%%%%%%%%%%%%%%

%%%%%%%%%%%%%%%%%%%%%%%%%
\begin{table}
\centering
\begin{minipage}{140mm}
 \caption{Parameters data of the CCSNe sample} 
\bigskip
\begin{tabular}{c c c c c c c c}
\hline \hline
SN &SN&Parent  &Distance & $A_{V}^{tot}$ & $M_{Ni}$ &$M_{oxy}$& References   \\
name &Type&galaxy & (Mpc) &  &$M_\odot$&$M_\odot$& \\
\hline
1987A&II&LMC&0.05 &0.6&0.075&$1.5-2$ & 1, 2\\
1988A&IIP&NGC 4579&22.95 &0.136&0.088 &$0.7-1.38$&7, 8\\
1988H&IIP&NGC 5878&28.47 &0.47&0.033&$0.79-1.1$& 8\\
1990E&IIP&NGC 1035&16.18 &1.2 & 0.043 &$0.94-1.3$& 10, 13\\
1991G&IIP&NGC 4088 &14.11 &0.065 &0.021&$0.36-0.49$ & 9\\
1992H&IIP&NGC 5377&29.07 &0.054 &0.123&$0.98-1.5$ & 14\\
1995ad&IIP&NGC 2139&23.52&0.112&$---$&$---$& 17\\
1997D&IIP&NGC 1536 &16.84 &0.07 &0.0065& $---$& 11, 12, 15\\
1999em&IIP&NGC 1637&8.8&0.31&0.027&$0.3-0.54$& 3, 4, 5 \\
2002hh&IIP&NGC 6946 &5.9 & 3.3 $\&$ 1.7 &0.07 &$0.78-1.05$ & 18\\
2004et&IIP&NGC6946&5.6 &1.27&0.06&$0.75-1.8$ & 19\\
1970G&IIL-P&NGC 5457 &7.2&0.4&0.051&0.88$-$1.18& 6, 16\\
1980K&IIL&NGC6946&5.6&1.5&0.046&0.57$-$0.76& 35, 36\\
1990I&Ib&NGC4650&39.3&0.4&0.11&0.7$-$1.35& 20\\
1996aq&Ib/c&NGC5584&24.32&0.124&$---$&0.4$-$1&17\\
1983I&Ic&NGC4051&22.08&0.043&0.15&0.43& 21\\
1983N&Ib&NGC5236&4.46&0.51&0.15&0.43& 21, 22\\
1990B&Ib/c&NGC4568&15.27&2.64&0.14&0.4& 23, 24, 25 \\
1984L&Ib&NGC991&23.44&0.32&0.37&0.58& 25, 26\\
1985F&Ib&NGC4618&7.3&0.0&0.1&0.3& 26\\
1987M&Ic&NGC 2715&22.7&1.3&0.26&0.4& 27, 28\\
1994I&Ic&NGC 5194&8.32&0.93&0.07&0.22& 29 \\
1993J&IIb&M81&3.64&0.6&0.1$-$0.14&0.5& 30, 31, 32 \\
1996N&Ib&NGC 1398&22&0.0&$---$&0.1$-$0.3& 30\\
1998bw&Ic&ESO 184-G82&35.16&0.2&$0.4-0.5$&$5-6$& 33\\
2002ap&Ic&M74&7.9&0.24&0.09&0.6& 34\\ 
\hline \hline 
\end{tabular} 
\end{minipage}
\\
\emph{{\rm \scriptsize \\ {\bf REFERENCES:}\\
 1- Arnett 1996; 2- Hirata et al. 
 1987; 3- Baron et al. 2000; 4- Elmhamdi et 
 al. 2003; 5- Hamuy et al. 2001; 6- Kirshner \& Kwan 1975; 
 7- Ruiz-lapuente et al. 1990; 8- 
 Turatto et al. 1993; 9- Blanton et al. 1995; 
 10- Benetti et al. 1994; 11- Turatto et al. 1998; 
 12- Benetti et al. 2001; 13- Schmidt et al. 1993;
 14- Clocchiatti et al. 1996; 15- Zampieri et al. 2003; 
 16- Barbon et al. 1973; 17- Based only on the present work;  
 18- Pozzo et al. 2006; 19- sahu et al. 2006; 
 20- Elmhamdi et al. 2004; 21- Shigeyama et al. 1990; 
 22- Clocchiatti et al. 1996c; 23- Clocchiatti et al. 2001; 
 24- Gomez $\&$ Lopez 2002; 25- Richardson et al. 2006; 26- Schlegel $\&$
 Kirshner 1989; 27- Filippenko et al. 1990; 28- Nomoto et al. 1990;
 29- Sauer et al. 2006; 30- Sollerman et al 1998; 31- Lewis et al. 1994;
 32- Barbon et al. 1995; 33- Maeda et al. 2006; 34- Foley et al. 2003;
 35- Buta 1982; 36- Uomoto $\&$ Kirshner 1986.\\
 \vspace{0.2truecm}
{\bf NOTE:} the net effect from the uncertainty sources, see text, indicate 
a variation in the calculated oxygen masses within $15-30 \%$. The estimated 
 iron mass is found to be less biased, especially for type II events. \\ } }
\normalsize
\end{table}

%%%%%%%%%%%%%%%%%%%%%%%%%%%%%%%

%%%%%%%%%%%%%%%%%%%%%%%%%


\begin{references}

\refitem{Anders E. $\&$ Grevesse N.}{1989}{Geochimica et cosmochimica Acta}{53}{ 197}
\refitem{Arnett W. D.}{1982}{ApJ}{253}{785}
\refitem{Barbon R., Ciatti F. $\&$ Rosino L.}{1973}{A$\&$A}{29}{57}
\refitem{Barbon R. et al.}{1995}{A$\&$AS}{110}{513}
\refitem{Baron E., Branch D, Hauschildt P. et al.}{2000}{ApJ}{545}{444}
\refitem{Benetti S., Cappellaro E., Turatto M. et al.}{1994}{A$\&$A}{285}{147}
\refitem{Benetti S.; Cappellaro E.; Danziger I. J. et al.}{1998}{MNRAS}{294}{448}
\refitem{Benetti S., Turatto M., Balberg S. et al.}{2001}{MNRAS}{322}{361}
\refitem{Blanton E. L., Schmidt B. P., Kirshner R. P. et al.}{1995}{AJ}{110}{2868}
\refitem{Buta, R. J.}{1982}{PASP}{94}{578}
\refitem{Cardelli J. A., Clayton G. C., Mathis J. S.}{1989}{ApJ}{345}{245}
\refitem{Castor J. I.}{1970}{MNRAS}{149}{111}
\refitem{Catchpole R. M., Whitelock P. A., Feast M. W. et al.}{1988}{MNRAS}{231}{75}
\refitem{Chugai N. N.}{1994}{ApJ}{428}{17}
\refitem{Chugai N. N.and Danziger I. J.}{2003}{AstL}{29}{649}
\refitem{Clocchiatti A. et al.}{1996a}{AJ}{111}{1286}
\refitem{Clocchiatti A. et al.}{1996b}{ApJ}{462}{462}
\refitem{Clocchiatti A.; Wheeler J. C.; Benetti S.; Frueh M.}{1996c}{ApJ}{459}{547}
\refitem{Clocchiatti A.; Suntzeff N. B.; Phillips M. M.; Filippenko A. V.; Turatto M. et al.}{2001}{ApJ}{553}{886}
\refitem{Danziger I. J.; Bouchet P.; Gouiffes C.; Lucy L. B.}{1991}{in ``Supernova 1987A and other supernovae'', ESO Conference and 
 Workshop Proceedings, Ed. by I. J. Danziger and Kurt Kjar}{}{217}
\refitem{de Kool M., Li H. $\&$ McCray R.}{1998}{ApJ}{503}{857}
\refitem{Elmhamdi A. et al.}{2003a}{MNRAS}{338}{939}
\refitem{Elmhamdi A.; Chugai N. N.; Danziger I. J.}{2003b}{A$\&$A}{404}{1077}
\refitem{Elmhamdi A., Danziger I. J., Cappellaro E. et al.}{2004}{A$\&$A}{426}{963}
\refitem{Elmhamdi A.; Danziger I. J.; Branch D. et al.}{2006}{A$\&$A}{450}{305}
\refitem{Elmhamdi A. et al.}{2011}{ApJ}{731}{129}
\refitem{Foley R. J.; Papenkova M. S.; Swift B. J. et al.}{2003}{PASP}{115}{1220}
\refitem{Filippenko A. V., Porter A. C. $\&$ Sargent W.}{1990}{AJ}{100}{1575}
\refitem{Filippenko A. V.}{1997}{ARA$\&$A}{35}{309}
\refitem{Fransson C.; Chevalier R. A.}{1987}{ApJ}{322}{15}
\refitem{Fransson C.; Chevalier R. A.}{1989}{ApJ}{343}{323}
\refitem{Fransson C.; Houck J.; Kozma C.}{1996}{in ``Supernovae and supernova remnants''. Proceedings of the IAU Colloquium 145;
 Ed. by Richard McCray and Zhenru Wang}{}{211}
\refitem{Fransson C.; Chevalier R. A.; Filippenko A. V.et al.}{2002}{ApJ}{572}{350}
\refitem{Gomez G. $\&$ Lopez R.}{2002}{AJ}{123}{328}
\refitem{Houck J. C. $\&$ Fransson C.}{1996}{ApJ}{456}{811}
\refitem{Hamuy M., Pinto P. A., Maza J. et al.}{2001}{ApJ}{558}{615}
\refitem{Hamuy M.}{2003}{ApJ}{582}{905}
\refitem{Jeffery D. J. $\&$ Branch D.}{1990}{in ``Supernovae'', Sixth Jerusalem Winte School for Theoretical Physics (Singapore: World Scientific); Ed. by
 J.C. Wheeler, T. Piran, $\&$ S. Weinberg}{}{149}
\refitem{Kirshner R. P. $\&$ Kwan J.}{1975}{ApJ}{197}{412}
\refitem{Leibundgut B., Kirshner R. P., Pinto P. A. et al.}{1991}{ApJ}{372}{531}\refitem{Lewis J. R., Walton N. A., Meikle W. P. S. et al.}{1994}{MNRAS}{266}{27}
\refitem{Li H. $\&$ McCray R.}{1992}{ApJ}{387}{309}
\refitem{Lucy L. B.; Danziger I. J.; Gouiffes C.; Bouchet P.}{1989}{in ``Structure and Dynamics of the Interstellar Medium'', Proceedings 
of IAU Colloq. 120. Ed. by Guillermo Tenorio-Tagle, Mariano Moles, 
 and Jorge Melnick.}{}{}
\refitem{Maeda K.; Nomoto K.; Mazzali P. A.; Deng J.}{2006}{ApJ}{640}{854}
\refitem{Matheson T. et al.}{2001}{AJ}{121}{1648}
\refitem{Maurer, J.I. et al.}{2010}{MNRAS}{402}{161}
\refitem{Mazzali, P.A. et al.}{2010}{MNRAS}{408}{87}
\refitem{McCray R.}{1996}{in ``Supernovae and supernova remnants''.
 Proceedings of the IAU Colloquium 145;
 Ed. by Richard McCray and Zhenru Wang}{}{223}
\refitem{Melendez J.; Barbuy B.; Spite F.}{2001}{ApJ}{556}{858}
\refitem{Menzies J. W.}{1991}{in ``Supernova 1987A and other supernovae'', 
 ESO Conference and Workshop Proceedings, Ed. by I. J. Danziger and 
 Kurt Kjar}{}{209}
\refitem{Nomoto K. et al.}{1984}{ApJ}{286}{644}
\refitem{Nomoto K.; Filippenko A. V.; Shigeyama T.}{1990}{A$\&$A}{240}{1}
\refitem{Nomoto K. et al.}{1997}{Nucl. Phys.}{616}{79}
\refitem{Pastorello A.; Turatto M.; Benetti S.; Cappellaro E.; Danziger I. J. et al.}{2002}{MNRAS}{333}{27}
\refitem{Pastorello A.; Aretxaga I.; Zampieri L.; Mucciarelli P.; Benetti S.}{2002}{ASPC}{342}{285}
\refitem{Patat F., Cappellaro E., Danziger J. et al.}{2001}{AJ}{555}{900}
\refitem{Popov D. V. }{1993}{ApJ}{414}{712}
\refitem{Pozzo M.; Meikle W. P. S.; Rayner J. T et al.}{2006}{MNRAS}{368}{1169}
\refitem{Qiu Y. et al.}{1999}{AJ}{117}{736}
\refitem{Richardson D.; Branch D.; Baron E.}{2006}{AJ}{131}{2233}
\refitem{Ruiz-Lapuente P., Canal R., Kidger M., Lopez R. et al.}{1990}{ApJ}{100}{782}
\refitem{Sahu D. K.; Anupama G. C.; Srividya S.; Muneer S.}{2006}{MNRAS}{372}{1315}
\refitem{Sauer D. N.; Mazzali P. A.; Deng J. et al.}{2006}{MNRAS}{369}{1939}
\refitem{Schlegel E. M. $\&$ Kirshner R. P.}{1989}{AJ}{98}{577}
\refitem{Schmidt B. P., Kirshner R. P., Schild R. et al.}{1993}{ApJ}{105}{2236}
\refitem{Shigeyama T.; Nomoto K.; Tsujimoto T.; Hashimoto M.}{1990}{ApJ}{361}{23}
\refitem{Sollerman J., Leibundgut B. $\&$ Spyromilio J.}{1998}{A$\&$A}{337}{207}\refitem{Spyromilio J. $\&$ Pinto P. A.}{1991}{in ``Supernova 1987A and other supernovae'', ESO Conference and Workshop Proceedings, Ed. by I. J. Danziger and 
 Kurt Kjar}{}{423}
\refitem{Suntzeff N. B. $\&$ Bouchet P.}{1991}{In ``Supernovae'', Ed. Woosley S. E., Springer-Vergal, New York}{}{12}
\refitem{Taubenberger, S. et al.}{2009}{MNRAS}{397}{677}
\refitem{Thielemann F. K., Nomoto K. $\&$ Hashimoto M.}{1996}{ApJ}{460}{408}
\refitem{Turatto M., Cappellaro E., Benetti S. $\&$ Danziger I. J.}{1993}{MNRAS}{265}{471}
\refitem{Turatto M., Mazzali P. A., Young T. R. et al.}{1998}{AJ}{498}{129}
\refitem{Uomoto A.}{1986}{ApJ}{310}{35}
\refitem{Uomoto A. $\&$ Kirshner R. P.}{1986}{ApJ}{308}{685}
\refitem{VandenBerg D. A.; Swenson F. J.; Rogers F. J.; Iglesias C. A.; Alexander D. R.}{2000}{ApJ}{532}{430}
\refitem{Woosley S.E. $\&$ Weaver T.A.}{1995}{ApJ}{101}{181}
\refitem{Zampieri L. et al.}{2003}{MNRAS}{338}{711}

\end{references}
\end{document}